# Observation of temporal topological boundary states of light in a momentum bandgap

Yudong Ren[1,2,3,4,#], Kangpeng Ye[1,#], Qiaolu Chen[1,2,3,4], Fujia Chen[1,2,3,4], Li Zhang[1,2,3,4], Yuang Pan[1,2,3,4], Wenhao Li[1,2,3,4], Xinrui Li[1,2,3,4], Lu Zhang[1,*], Hongsheng Chen[1,2,3,4,*], Yihao Yang[1,2,3,4,*]

[1] State Key Laboratory of Extreme Photonics and Instrumentation, ZJU-Hangzhou Global Scientific and Technological Innovation Center, Zhejiang University, Hangzhou 310027, China.

[2] International Joint Innovation Center, The Electromagnetics Academy at Zhejiang University, Zhejiang University, Haining 314400, China

[3] Key Lab. of Advanced Micro/Nano Electronic Devices & Smart Systems of Zhejiang, Jinhua Institute of Zhejiang University, Zhejiang University, Jinhua 321099, China

[4] Shaoxing Institute of Zhejiang University, Zhejiang University, Shaoxing 312000, China

#These authors contributed equally to this work.

*Correspondence to: (Y. Y.) yangyihao@zju.edu.cn; (L. Z.) zhanglu1993@zju.edu.cn; (H. C.) hansomchen@zju.edu.cn

## ABSTRACT

Topological phases have prevailed across diverse disciplines, spanning electronics, photonics, and acoustics. Hitherto, the understanding of these phases has centred on energy (frequency) bandstructures, showcasing topological boundary states at spatial interfaces. Recent strides have uncovered a unique category of bandstructures characterized by gaps in momentum, referred to as momentum bandgaps or *k* gaps, notably driven by breakthroughs in photonic time crystals[1,2]. This discovery hints at abundant topological phases defined within momentum bands, alongside a wealth of topological boundary states in the time domain. Here, we report the first experimental observation of *k*-gap topology in a large-scale optical temporal synthetic lattice, manifesting as temporal topological boundary states. These boundary states are uniquely situated at temporal interfaces between two subsystems with distinct *k*-gap topology. Counterintuitively, despite the exponential amplification of *k*- **gap modes within both subsystems, these topological boundary states exhibit decay in both temporal directions. Our findings mark a significant pathway for delving into $k$ gaps, temporal topological states, and time-varying physics.**

The investigation of topological phases has traditionally been rooted in electronic systems[3], while analogous behaviours have been identified across various physical domains, including but not limited to photonic[4], acoustic[5], and mechanical systems[6]. Currently, the understanding of topological phases has primarily centred on the global behaviours of wavefunctions within the energy (frequency) bands of materials or artificial structures, spanning diverse space groups and symmetries in real or synthetic dimensions[7,8]. The development of topological phases within frequency bands has led to the discovery of numerous captivating physical phenomena, as exemplified by the topological boundary states that immune to scattering from disorders or imperfections, and promising applications in quantum computing, lasers, and communications[9–14].

Very recently, it has been noted that the bandstructures can be gapped in momentum (or Bloch momentum), with the gaps known as momentum bandgaps or $k$ gaps[1,15–19]. This surge in interest in $k$ gaps is largely stimulated by the advent of photonic time crystals, which modulate permittivity periodically and abruptly in time[1,15,16,20]. The $k$ gaps differ from the frequency bandgaps in fundamental aspects. Firstly, unlike the frequency bandgaps that emerge from spatial reflection and refraction in spatially periodic materials[21], the $k$ gaps typically arise from interference between time-reflected[22–24] and time-refracted waves in temporally periodic materials. Secondly, while frequency-gap modes always decay in space and cannot propagate, $k$-gap modes exhibit exponential growth or decay in time while retaining the ability to propagate in space[1,25]. These $k$ gaps manifest diverse physical phenomena[1,2,19,25–27], such as enhanced light-electron interactions[25] and superluminal solitons[27], offering potential for various device applications, notably non-resonant tuneable lasers[2].

Much akin to frequency bandgaps, $k$ gaps also exhibit topological properties defined by momentum bands lower than the $k$ gaps[1,28]. However, the fundamental discrepancies between these two categories of bandgaps yield distinctive topological phenomena. For instance, the topological boundary state within the frequency bandgap emerges at a spatial interface (see **Fig. 1a**) where frequency is conserved but momentum is not, while the topological boundary state within the $k$ gap arises at a temporal interface (see **Fig. 1b**) conserving momentum but not frequency. Intriguingly, the spatial topological boundary state confines itself to an interface between two

"insulators," whereas its temporal counterpart localizes at an interface between two media supporting exponentially amplifying *k*-gap modes. Moreover, the existence of temporal topological boundary states correlates with the principle of causality. Gaining insights into *k*-gap topology holds immense potential for exploring and harnessing various topological phases defined within momentum bands and temporal topological effects across quantum and classical systems. To date, the *k*-gap topology has been theoretically proposed in photonic time crystals[1]. However, its experimental demonstration remains elusive, largely due to the difficulty in periodically modulating refractive indices of photonic materials within extremely short timescales and comparable amplitudes to the materials, particularly at optical frequencies.

Here, we report the experimental observation of temporal topological boundary states stemming from *k*-gap topology of light. Our experimental framework is based on an optical system comprising two fibre loops[29–34], which is mapped into a large-scale temporal synthetic lattice featuring a Bloch-momentum bandstructure inclusive of a complete *k* gap. Using this setup, we directly observe the emergence of topological boundary states at temporal interfaces between two subsystems displaying distinct *k*-gap topology. Intriguingly, these boundary states demonstrate a counterintuitive behaviour, decaying away from the interface in both temporal directions, despite both subsystems supporting exponential growth of *k*-gap modes. Furthermore, we observe the distinctive behaviour of optical pulses experiencing time-refraction and time-reflection within the *k* gap and the momentum bands. Specifically, when traversing a temporal slab, pulses within the momentum band bifurcate into four distinct pulses, whereas their *k*-gap counterparts diverge into two amplified pulses.

In our experiment, optical pulses propagate in two fibre loops with slightly different lengths connected by a variable optical coupler (VOC), as shown in **Fig. 1c**. This coupler facilitates a tuneable coupling ratio $\beta$, controlled by an electric signal sourced from an arbitrary waveform generator (AWG). A 50 ns rectangular-shaped pulse is injected into the longer loop via a 50:50 optical coupler (OC). The pulse is generated by a 1550 nm distributed-feedback laser, with its output beam modulated by an acousto-optic modulator (AOM). Utilizing a Mach-Zehnder modulator (MZM), we can manipulate the signal amplitude within the loops. Besides, a phase modulator (PM) is integrated into the loops to manipulate the phase of the optical signal, which is used to generate a pulse with predesigned Bloch momenta (**see supplementary information**). The optical losses in both loops are compensated by erbium-doped fibre amplifiers (EDFAs). As the optical pulses propagate in the system, the power in each loop is monitored using photodetectors (PDs).

In our fibre-loop system, pulses in the shorter loop experience a temporal advancement, whereas those within the longer loop undergo a delay. Such a pulse delay or advancement in the time domain, within a round-trip, can be conceptually analogous to a distance in the space dimension. Concurrently, the count of round-trips inherently offers a temporal degree of freedom, which can be viewed as a time dimension. As a result, this fibre-loop system can be equivalently mapped into a spatiotemporal lattice network, as illustrated in **Fig. 1d**. The pulse evolution in the lattice is governed by the following equations that describe light dynamics in a temporal unit comprising two successive time steps,

$$u_n^{m+1} = [\cos(\beta)u_{n+1}^m + i\sin(\beta)v_{n+1}^m]e^{\gamma(m)},$$
$$v_n^{m+1} = i\sin(\beta)u_{n-1}^m + \cos(\beta)v_{n-1}^m \quad (1)$$

where $u_n^m$ denotes the amplitude at lattice position $n$ and time step $m$ on left-moving paths, and $v_n^m$ denotes the corresponding amplitude on right-moving paths. $\gamma(m)$ is a temporally amplitude-modulated signal applied to the left-moving paths, satisfying

$$\gamma(m) = \begin{cases} +\gamma_0 & \text{for} \quad \mod(m;2) = 1 \\ -\gamma_0 & \text{for} \quad \mod(m;2) = 0 \end{cases}. \quad (2)$$

By employing the Floquet-Bloch ansatz, we deduce the system's bandstructure, showcased in **Fig. 1e**. The relation between quasienergy, $\theta$ (or the longitudinal propagation constant) and Bloch momentum (or the transverse wavenumber) along the 1D lattice, $Q$ is given by

$$\cos(\theta) = \cos^2(\beta)\cos(Q) - \sin^2(\beta)\cosh(\gamma_0). \quad (3)$$

Interestingly, the above bandstructure is gapped in the momentum, around $Q = \pi$, signifying an absence of real-value solutions for quasienergy. We note that our system is "spatially" uniform yet temporally modulated, which preserves the "spatial" translational symmetry but breaks the temporal translational symmetry. At each temporal interface, the Bloch momentum $Q$ (not the quasienergy $\theta$) is conserved, which is thus a good quantum number. This is fundamentally different from spatially modulated yet stationary systems, such as spatial photonic crystals, in which frequency is conserved while momentum is not. Besides, throughout our experiments (including those of topological boundary states), the "spatial" (temporal) translational symmetry is preserved (broken). In view of the above facts, we consider the bandgap here as a complete $k$ gap. In some sense, our temporal synthetic lattice resembles the photonic time crystal, despite the fact that the modes in our lattice extract the energy from amplifiers and the latter from the modulation of refraction indices; for example, modes propagating in two loops can be mapped respectively to the forward- and backward-propagating modes in the photonic time crystal (**see supplementary materials and Fig. S6**).

We now perform experiments to investigate the time-refraction and time-reflection behaviour of light in the momentum bands or within the $k$ gap, respectively. In the first case, the lattice framework has triple time periods, mimicking a temporal slab (**Fig. 2a, b**). To better illustrate the light behaviour solely attributed to the bands, all three periods feature no $k$ gaps, and the first and third time periods have distinct bandstructures from the second one (**Fig. 2e**). At time step $m = 0$, a Gaussian-shaped wave packet from the upper band is excited, which carries an initial momentum $Q = 0.95\pi$ and propagates forward. Upon reaching the first temporal interface at $m = 21$, the momentum conservation principle dictates that the pulse couples to two distinct sets of Bloch modes in the second period: a time-refracted pulse propagating forward and its time-reversed counterpart propagating backwards. At the subsequent temporal interface at $m = 41$, a similar bifurcation occurs, yielding a total of four pulses out from the temporal slab (**Fig. 2b, c**). Interestingly, the pulse propagation in the temporal slab markedly differs from that in a spatial counterpart, which undergoes infinite times of refraction and reflection at two interfaces, resulting in superpositions of refracted and reflected waves in the spatial slab. As the bands are purely real, energy is conserved in this scenario, as shown in **Fig. 2d**.

In the second case, there are also triple time periods (**Fig. 2f, g**), however, with the second time period featuring a $k$ gap (**Fig. 2j**). Similarly, at $m = 0$, a Gaussian-shaped wave packet from the upper band is excited, which carries an initial momentum $Q = 0.95\pi$ inside the $k$ gap. Upon reaching the first interface ($m = 21$), the pulse bifurcates into two modes with zero group velocity (**Fig. 2g**), meaning they are "frozen" in space. These modes either increase or decrease exponentially in the second period; the total energy of the modes exhibits exponential growth (**Fig. 2i**), in contrast to the conserved energy in the first case. At the second temporal interface, the mode with amplifying energy dominates, further splitting into time-refracted and time-reflected modes (**Fig. 2g, h**). Thus, only two pulses are observed after the pulse traverses the temporal slab, which markedly differs from the first case.

Next, we perform experiments to study the temporal topological boundary states at the interface between two lattices with distinct *k*-gap topology (**Fig. 3a**). The *k*-gap topology can be understood from an effective momentum operator near $\theta = \pi$ (**see supplementary information**)

$$H_{\delta Q}(\delta\theta) = -m_{\gamma_0}\sigma_x + \frac{\delta\theta}{\alpha}\sigma_z \qquad (4)$$

where $\delta Q$ is the Bloch momentum deviated from $Q = \pi$, $\delta\theta$ is the quasienergy deviated from $\theta = \pi$, and $m_{\gamma_0} = \sinh(\gamma_0)$. $\alpha = \sqrt{2}/2$, corresponding to the group velocity at $\delta Q = 0$ in the lattice without gain and loss. Interestingly, Eq. (4) is analogue to the well-known 1D massive Dirac Hamiltonian[35]; Here, however, the eigenvalue is momentum rather than energy as in the usual Hamiltonian. The bandstructure defined by Eq. (4) can be viewed as a momentum bandstructure. Besides, when $\gamma_0$ switches from negative to positive, a momentum band inversion occurs around $\theta = \pi$, where two mutually orthogonal states exchange their positions (**Fig. 3b**). This is similar to the band inversion of spatial topological photonic crystals[28]. Similar to the Jackiw-Rebbi solution at a mass domain wall, when placing two media together with opposite mass terms $m_{\gamma_0}$, there exists a zero-momentum topological state localized at the temporal boundary (**see supplementary information**).

Owing to the disconnected parameter path in quasienergy, the Zak phase of the momentum band is not well-defined. Therefore, we compare the local geometric phases[36] between lattices with opposite $m_{\gamma_0}$. To do so, we perform parallel transport on the band lower than the *k* gap ($\delta Q < 0$), which yields a vanishing Berry connection everywhere on the momentum band. We choose the eigenstate at a low quasienergy limit, $[1,0]^T$, as an initial state with no phase difference. For the final state at a high quasienergy limit $[0,1]^T$, the phase is 0 for a positive $m_{\gamma_0}$, in contrast to $\pi$ for a negative $m_{\gamma_0}$ (**see supplementary information and Fig. S4**). This implies the distinct *k*-gap topology determined by the sign of $m_{\gamma_0}$.

According to topological physics, when placing together two subsystems with topologically different bandgaps, topological boundary states are formed at the interfaces. In our case, topological boundary states should appear at a temporal interface between subsystems with distinct *k*-gap topology. For experimental verification, a step-like temporal modulation $\gamma_0$ is applied (as depicted in **Fig. 3c**), abruptly altering the system's *k*-gap topology, creating a topological temporal interface at *m* = 30 (**Fig. 3c**). At *m* = 0, a Gaussian-shaped wave packet with a momentum residing in the *k* gap is excited. Before reaching the temporal interface, the pulse energy increases exponentially, as demonstrated in the previous section. After passing the temporal interface, the energy exhibits an exponential decrease, forming an energy peak localized at the interface (**Fig. 3d-f**). Interestingly, if these two time periods are considered separately, the pulse propagating in each of them should increase exponentially as time progresses. However, the energy decays in both temporal directions at the interface, which is completely counterintuitive. Upon comparing the characteristics of our temporal topological boundary states with the spatial equivalents in photonic crystals, striking differences emerge: In spatial scenarios, the topological edge states are confined around the interface between two regions, supporting only evanescent waves that decay in space. Conversely, in temporal scenarios like ours, the topological boundary states reside between two regions, each of which supporting modes growing exponentially in time and being able to propagate in space.

Further, topological temporal boundary states exhibit marked robustness against disorders akin to their spatial counterparts. To verify the robustness, we first perturb the temporal interface by adding several temporal layers with random $\gamma_0 \in [-0.23, 0.23]$ around the interface (**Fig. 3g**). Notably, we can identify a clear energy peak localized around the interface (**Fig. 3h-j**). We then perturb the temporal bulk by introducing spatially uniform disorders $\delta\gamma_0 \in [-\gamma_0/5, \gamma_0/5]$ (as illustrated in **Fig. 3k**) into the temporal bulk across all lattice sites. The energy peak at the topological interface can still be spotted (**Fig. 3l-n**). To conclude, the persistence of temporal topological interface states remains evident whether the disorder is introduced at the boundary or permeates the bulk. These results indicate the intrinsic robustness of these states against considerable disorder, highlighting their topological protection—a hallmark of topological systems.

To summarize, we experimentally confirm the topological properties of the *k* gap by directly observing temporal topological boundary states in a large-scale optical temporal synthetic lattice. Moreover, the propagation dynamics of the *k*-gap modes through a temporal slab are observed, particularly elucidating their time-refraction and time-reflection behaviour at temporal boundaries. Those phenomena are otherwise extremely challenging (if not impossible) to observe in real optical materials. Our work thus establishes an ideal platform using light to explore the unconventional physics associated with *k* gaps, encompassing wave dynamics and topological attributes. While our study predominantly focuses on the temporal dimension, an intriguing avenue for future exploration involves incorporating one or more spatial dimensions to form topological phases within a comprehensive space-time framework. Additionally, there exists potential for incorporating nonlinearity[30] and disorder[33] into our fibre-loop systems, leading to an intriguing synergy between *k*-gap topology, nonlinearity, and disorder.

## Acknowledgments

The work at Zhejiang University sponsored by the Key Research and Development Program of the Ministry of Science and Technology under Grants 2022YFA1405200 (Y.Y.), No.2022YFA1404704 (H.C.), 2022YFA1404902 (H.C.), and 2022YFA1404900 (Y.Y.), the National Natural Science Foundation of China (NNSFC) under Grants No. 62175215 (Y.Y.), and No.61975176 (H.C.), the Key Research and Development Program of Zhejiang Province under Grant No.2022C01036 (H.C.), the Fundamental Research Funds for the Central Universities (2021FZZX001-19) (Y.Y.), and the Excellent Young Scientists Fund Program (Overseas) of China (Y.Y.).

## Authors Contributions

Y.Y. initiated the idea. Y.Y. and Y.R. designed the experiment, K.Y. and Y.R. carried out the experiment with the assistance from Y.Y. and L.Z.. Y.R. and K.Y. analysed the data. Y.R. and Y.Y. performed the simulations. Y.R. and Y.Y. did the theoretical analysis. Y.R., Y.Y. wrote the paper. Y.Y., H.C., L.Z. supervised the project. All authors participated in discussions and reviewed the paper.

## Competing Interests

The authors declare no competing interests.

## Data availability

The data that support the findings of this study are available from the corresponding author upon reasonable request.

# References

1. Lustig, E., Sharabi, Y. & Segev, M. Topological aspects of photonic time crystals. *Optica* **5**, 1390 (2018).
2. Lyubarov, M. et al. Amplified emission and lasing in photonic time crystals. Science 377, 425–428 (2022).
3. Hasan, M. Z. & Kane, C. L. Colloquium: Topological insulators. *Rev. Mod. Phys.* **82**, 3045–3067 (2010).
4. Ozawa, T. *et al.* Topological photonics. *Rev. Mod. Phys.* **91**, 015006 (2019).
5. Xue, H., Yang, Y. & Zhang, B. Topological acoustics. *Nat. Rev. Mater.* **7**, 974–990 (2022).
6. Ma, G., Xiao, M. & Chan, C. T. Topological phases in acoustic and mechanical systems. *Nat Rev Phys* **1**, 281–294 (2019).
7. Wen, X.-G. *Colloquium*: Zoo of quantum-topological phases of matter. *Rev. Mod. Phys.* **89**, 041004 (2017).
8. Ozawa, T. & Price, H. M. Topological quantum matter in synthetic dimensions. *Nat Rev Phys* **1**, 349–357 (2019).
9. Bahari, B. *et al.* Nonreciprocal lasing in topological cavities of arbitrary geometries. *Science* **358**, 636–640 (2017).
10. Harari, G. *et al.* Topological insulator laser: Theory. *Science* **359**, eaar4003 (2018).
11. Bandres, M. A. *et al.* Topological insulator laser: Experiments. *Science* **359**, eaar4005 (2018).
12. Mikhail I. Shalaev, Wiktor Walasik, Alexander Tsukernik, Yun Xu, & Natalia M. Litchinitser. Robust topologically protected transport in photonic crystals at telecommunication wavelengths. *Nat. Nanotech.* **14**, 31–34 (2019).
13. He, X.-T. *et al.* A silicon-on-insulator slab for topological valley transport. *Nat. Commun.* **10**, 872 (2019).
14. Yang, Y. *et al.* Terahertz topological photonics for on-chip communication. *Nat. Photon.* **14**, 446–451 (2020).
15. Biancalana, F., Amann, A., Uskov, A. V. & O'Reilly, E. P. Dynamics of light propagation in spatiotemporal dielectric structures. *Phys. Rev. E* **75**, 046607 (2007).
16. Zurita-Sánchez, J. R., Halevi, P. & Cervantes-González, J. C. Reflection and transmission of a wave incident on a slab with a time-periodic dielectric function $\epsilon(t)$. *Phys. Rev. A* **79**, 053821 (2009).
17. Reyes-Ayona, J. R. & Halevi, P. Observation of genuine wave vector ($k$ or $\beta$) gap in a dynamic transmission line and temporal photonic crystals. *Appl. Phys. Lett.* **107**, 074101 (2015).
18. Li, M.-W., Liu, J.-W., Chen, W.-J. & Dong, J.-W. Topological momentum gap in PT-symmetric photonic crystals. Preprint at https://arxiv.org/abs/2306.09627 (2023).
19. Wang, X. *et al.* Metasurface-based realization of photonic time crystals. *Science Advances* **9**, eadg7541 (2023).
20. Lustig, E. *et al.* Photonic time-crystals - fundamental concepts [Invited]. *Opt. Express* **31**, 9165 (2023).
21. *Photonic crystals: molding the flow of light*. (Princeton University Press, 2008).
22. Moussa, H. *et al.* Observation of temporal reflection and broadband frequency translation at photonic time interfaces. *Nat. Phys.* **19**, 863–868 (2023).
23. Galiffi, E. *et al.* Broadband coherent wave control through photonic collisions at time interfaces. *Nat. Phys.* https://doi.org/10.1038/s41567-023-02165-6 (2023).
24. Dong, Z. *et al.* Quantum time reflection and refraction of ultracold atoms. *Nat. Photon.* https://doi.org/10.1038/s41566-023-01290-1 (2023).

25. Dikopoltsev, A. *et al.* Light emission by free electrons in photonic time-crystals. *Proc. Natl. Acad. Sci. U.S.A.* **119**, e2119705119 (2022).
26. Sharabi, Y., Lustig, E. & Segev, M. Disordered Photonic Time Crystals. *Phys. Rev. Lett.* **126**, 163902 (2021).
27. Pan, Y., Cohen, M.-I. & Segev, M. Superluminal k-Gap Solitons in Nonlinear Photonic Time Crystals. *Phys. Rev. Lett.* **130**, 233801 (2023).
28. Xiao, M., Zhang, Z. Q. & Chan, C. T. Surface Impedance and Bulk Band Geometric Phases in One-Dimensional Systems. *Phys. Rev. X* **4**, 021017 (2014).
29. Regensburger, A. *et al.* Parity–time synthetic photonic lattices. *Nature* **488**, 167–171 (2012).
30. Wimmer, M. *et al.* Observation of optical solitons in PT-symmetric lattices. *Nat. Commun.* **6**, 7782 (2015).
31. Wimmer, M., Price, H. M., Carusotto, I. & Peschel, U. Experimental measurement of the Berry curvature from anomalous transport. *Nature Phys* **13**, 545–550 (2017).
32. Weidemann, S. *et al.* Topological funneling of light. *Science* **368**, 311–314 (2020).
33. Weidemann, S., Kremer, M., Longhi, S. & Szameit, A. Coexistence of dynamical delocalization and spectral localization through stochastic dissipation. *Nat. Photon.* **15**, 576–581 (2021).
34. Weidemann, S., Kremer, M., Longhi, S. & Szameit, A. Topological triple phase transition in non-Hermitian Floquet quasicrystals. *Nature* **601**, 354–359 (2022).
35. Shen, S.-Q. *Topological Insulators: Dirac Equation in Condensed Matters*. (Springer Berlin Heidelberg, 2012).
36. Resta, R. Manifestations of Berry's phase in molecules and condensed matter. *J. Phys.: Condens. Matter* **12**, R107–R143 (2000).

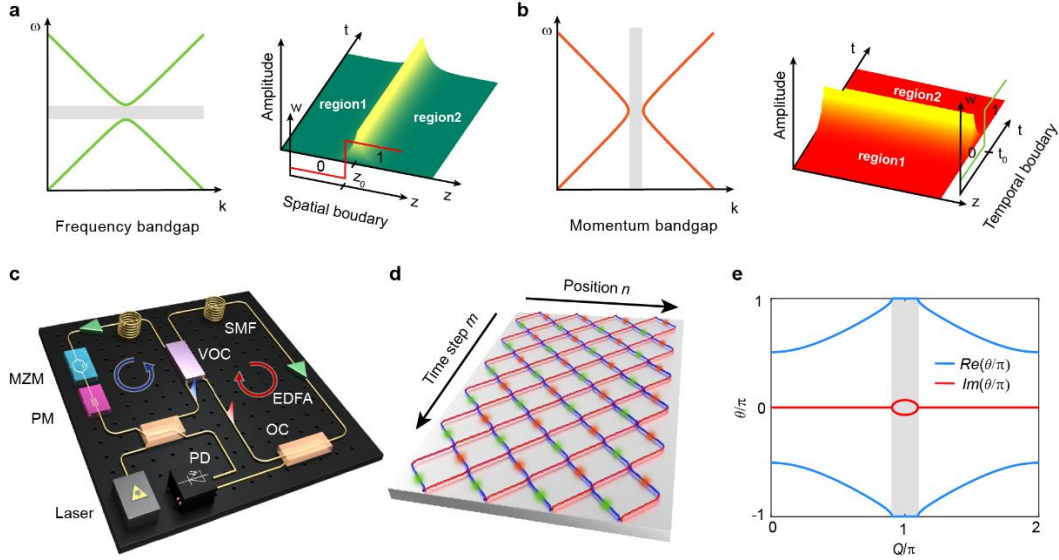

**Fig. 1 Topological boundary state in the *k* gap and the experimental setup. a**, Topological boundary states in a frequency bandgap. Topological boundary states are localized at a spatial interface between two insulators with topologically distinct frequency bandgaps, with the frequency conserved. Topological number *w* switches in the spatial dimension. **b**, Topological boundary states in a *k* gap. Topological boundary states are localized at a temporal interface between two media with topologically distinct *k* gaps, with the momentum conserved. Topological number *w* switches in the temporal dimension. **c**, Schematic diagram of the experimental setup. The system consists of two fibre loops with different lengths. PM: phase modulator; MZM: Mach-Zehnder modulators; OC: optical coupler; VOC: variable optical coupler; SMF: single-mode fibre; PD: photodetector; EDFA: erbium-doped fibre amplifier. **d**, Temporal synthetic lattice mapped from the setup in (**c**). The orange and green shadows represent gain ($+\gamma_0$) and loss ($-\gamma_0$), respectively, controlled by MZM. **e**, Complex bandstructure of the lattice when $\gamma_0 = 0.3$. A complete *k* gap opens near $Q = \pi$.

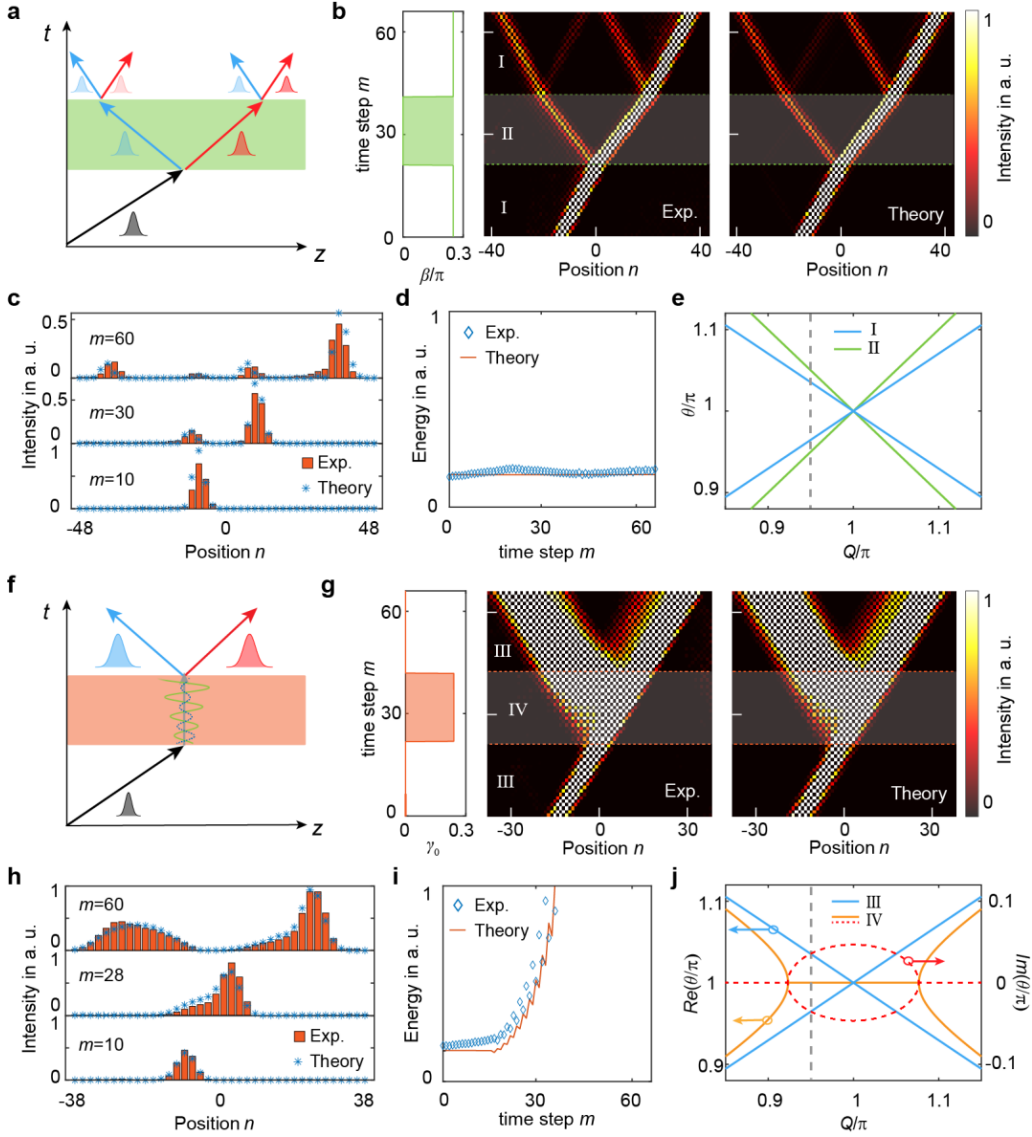

**Fig. 2. Pulse propagation in the *k* gap. a**, Schematic diagram of the time refraction and time reflection behaviour in the band. A pulse undergoes splitting event at each temporal interface. **b**, Parameter $\beta$ as a function of time (left panel), the measured pulse propagation (middle panel), and the calculated result (right panel). In the first and third time periods (from $m = 0$ to 21 and from $m = 41$ to 66), $\gamma_0 = 0$ and $\beta = \pi/4$, with the bandstructure corresponding to the blue line in (**e**). In the second time period (from $m = 21$ to 41), $\gamma_0 = 0$ and $\beta = 0$, with the bandstructure corresponding to the green line in (**e**). **c**, Pulse intensity distributions at $m = 10$, 30, and 60, respectively. **d**, Total energy of the pulses. **e**, Bandstructures in different time periods. The Grey dashed line indicates the conservation of Bloch momentum at the temporal interface. **f**, Schematic diagram of the time

refraction and time reflection behaviour in the $k$ gap. **g**, Parameter $\gamma_0$ as a function of time (left panel), the measured pulse propagation (middle panel), and the calculated result (right panel). In the first and third time periods (from $m = 0$ to 21 and from $m = 41$ to 66), $\gamma_0 = 0$ and $\beta = \pi/4$, with the bandstructure corresponding to the blue line in (**j**). In the second time period (from $m = 21$ to 41), $\gamma_0 = 0.23$ and $\beta = \pi/4$, with the bandstructure corresponding to the orange (real part) and red (imaginary part) lines in (**j**). **h**, Pulse intensity distributions at $m = 10$, 28, and 60, respectively. **i**, Total energy of the pulses. **j**, Bandstructures in different time periods. The real and imaginary parts are represented by the solid and dashed lines, respectively.

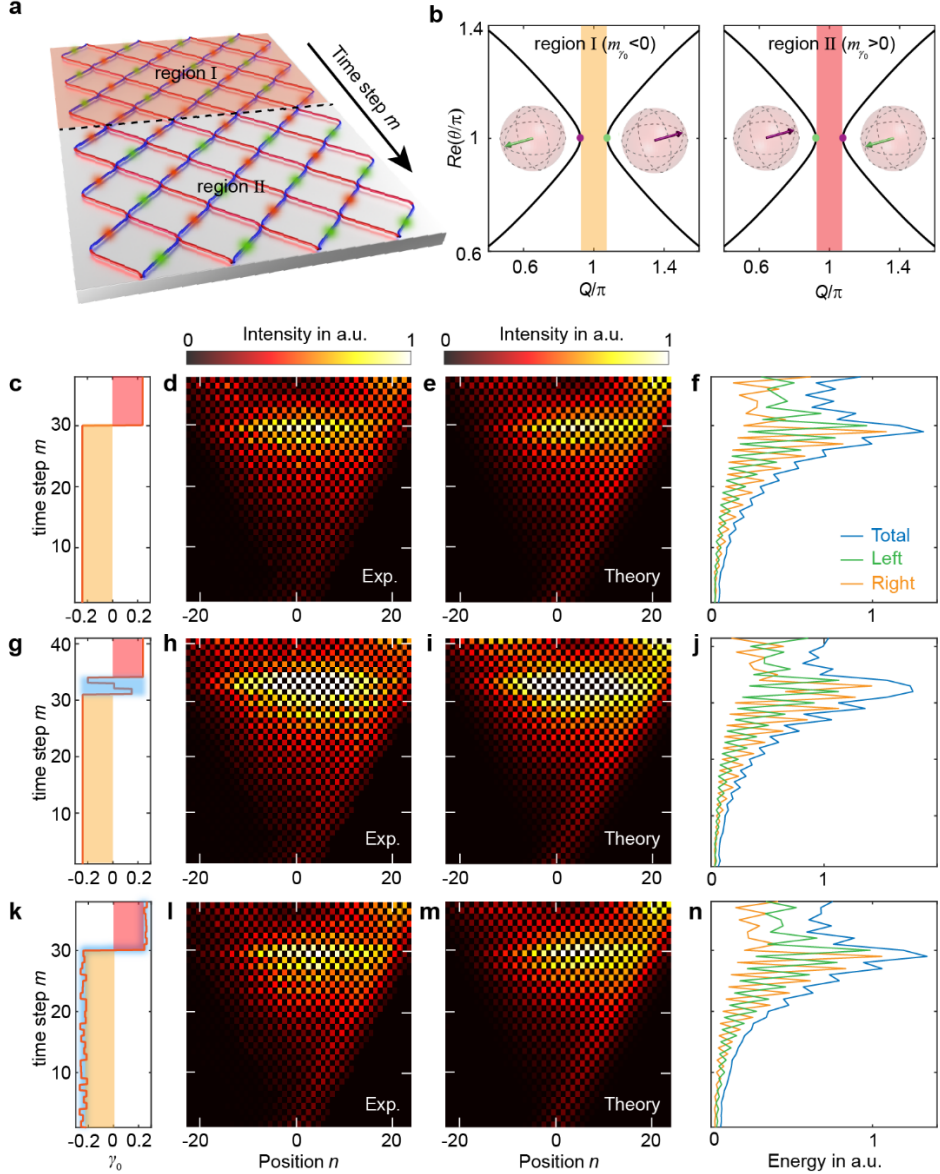

**Fig. 3 Robust temporal topological boundary states from *k*-gap topology. a**, Schematic diagram of a topological temporal interface in the synthetic lattice. **b**, Bandstructures of two lattices with distinct *k*-gap topology. The insets show two eigenvectors at $\theta = \pi$. In the region I ($\gamma_0 > 0$), the eigenvector $[\bar{U}, \bar{V}]^T$ at the left (right) band is $[-1/\sqrt{2}, 1/\sqrt{2}]^T$ ($[1/\sqrt{2}, 1/\sqrt{2}]^T$), while in the region II ($\gamma_0 < 0$), the eigenvectors are inverted for two bands. **c-f**, Experimental observation of temporal topological edge states. Parameter $\gamma_0$ as a function of time (**c**), the measured pulse propagation (**d**), and the calculated result (**e**). Measured light energy (**f**). In the region I ($m < 30$), $\gamma_0 = 0.23$, In the region II ($m > 30$), $\gamma_0 = 0.23$. In the entire process, $\beta = \pi/4$. **g**-

**n**, Temporal topological boundary with disorder around the boundary (**g-j**) and in the bulk (**k-n**). The blue shadow area represents the time duration when the disorder is applied.